\documentclass{article}
\usepackage{spconf,amsmath,graphicx,hyperref}
\usepackage{booktabs}
\usepackage{tabularx}  %
\usepackage[capitalize]{cleveref}

\title{High-Fidelity Speech Enhancement via Discrete Audio Tokens}
\name{
    Luca A. Lanzendörfer \qquad Frédéric Berdoz \qquad Antonis Asonitis \qquad Roger Wattenhofer
    \address{ETH Zurich}
}
\begin{document}
\maketitle
\begin{abstract}
Recent autoregressive transformer-based speech enhancement (SE) methods have shown promising results by leveraging advanced semantic understanding and contextual modeling of speech. However, these approaches often rely on complex multi-stage pipelines and low sampling rate codecs, limiting them to narrow and task-specific speech enhancement. In this work, we introduce DAC-SE1, a simplified language model-based SE framework leveraging discrete high-resolution audio representations; DAC-SE1 preserves fine-grained acoustic details while maintaining semantic coherence. Our experiments show that DAC-SE1 surpasses state-of-the-art autoregressive SE methods on both objective perceptual metrics and in a MUSHRA human evaluation. We release our codebase and model checkpoints to support further research in scalable, unified, and high-quality speech enhancement.\footnote{\url{https://github.com/ETH-DISCO/DAC-SE1}}
\end{abstract}

\section{Introduction}

Scaling laws have transformed machine learning across multiple domains, from natural language processing~\cite{brown2020language, chowdhery2023palm}
and computer vision~\cite{dosovitskiy2020image, kirillov2023segment} 
to speech and audio generation~\cite{borsos2023audiolm, radford2023robust}.
In particular, large autoregressive transformer models (LLMs) trained on discrete audio representations have achieved remarkable performance in text-to-speech~\cite{ye2025llasa}, audio synthesis~\cite{borsos2023audiolm, copet2023simple}, and speech understanding~\cite{team2023gemini}, demonstrating that model size and data scale can naturally improve both fidelity and generalization.
Despite these advances, speech enhancement (SE) remains dominated by models that either operate in the time domain or by models using conditional architectures~\cite{kang2025llase, wang2024selm}. Time domain models such as Conv-TasNet~\cite{luo2019conv}, Demucs~\cite{defossez2020real}, and DCCRN~\cite{hu2020dccrn} or LM-based frameworks operate on either low sampling rate codecs or by using multi-stage architectures. Although effective to some extent, these approaches introduce architectural modifications that can hinder scalability and high-fidelity reconstruction. In this work, we investigate whether high-quality speech enhancement can be achieved solely through scaling laws in data and compute, without domain-specific adaptations.
\begin{figure}[t!]
    \centering
    \includegraphics[width=\columnwidth]{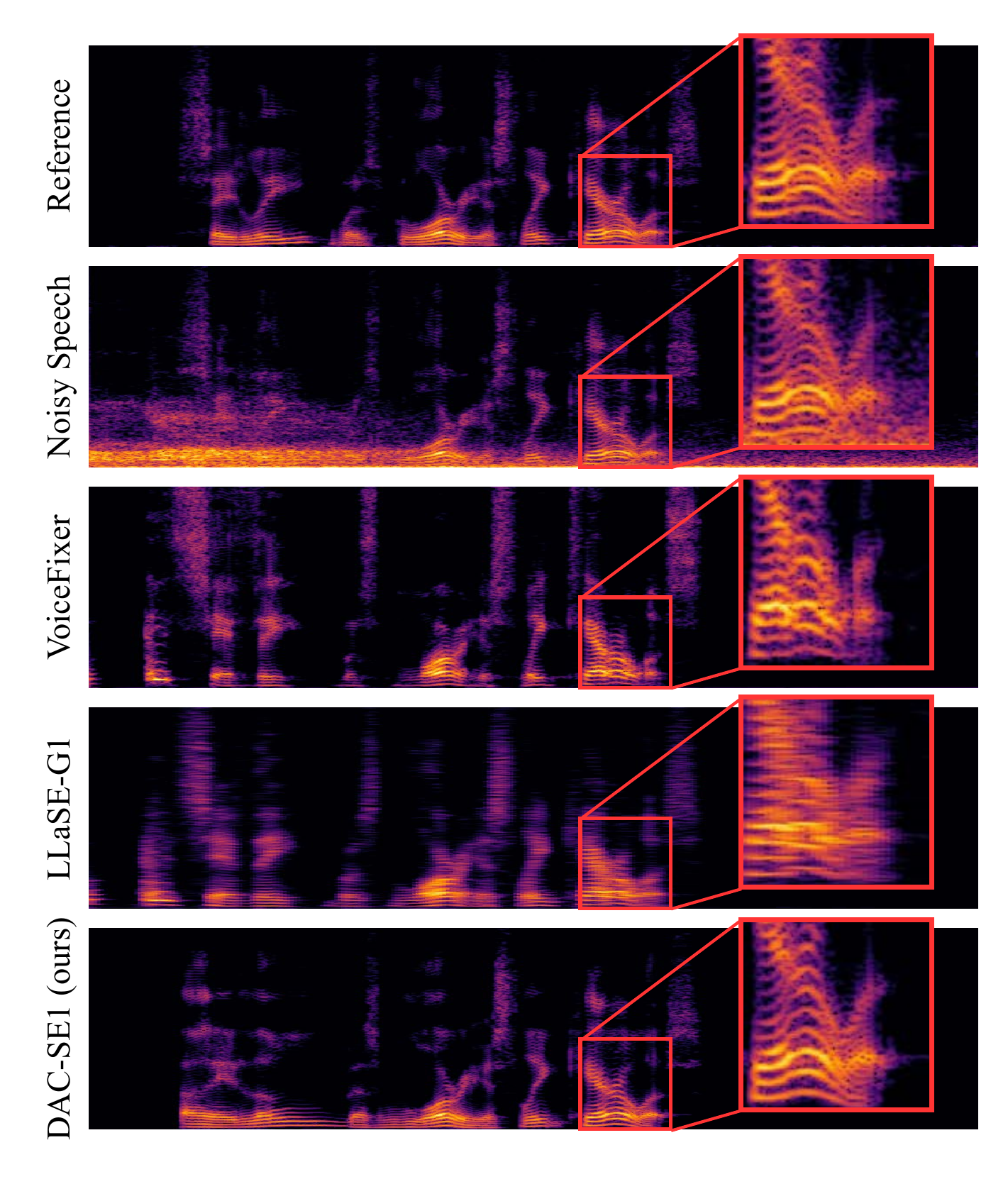}
    \caption{Qualitative comparison on log-mel spectrograms between our proposed method (DAC-SE1) and previous autoregressive speech enhanecment methods. DAC-SE1 is able to clean the signal without hallucinating artifacts or spectral distortion.}
    \label{fig:qual}
\end{figure}
\begin{figure*}[t]
    \centering
    \includegraphics[width=\textwidth]{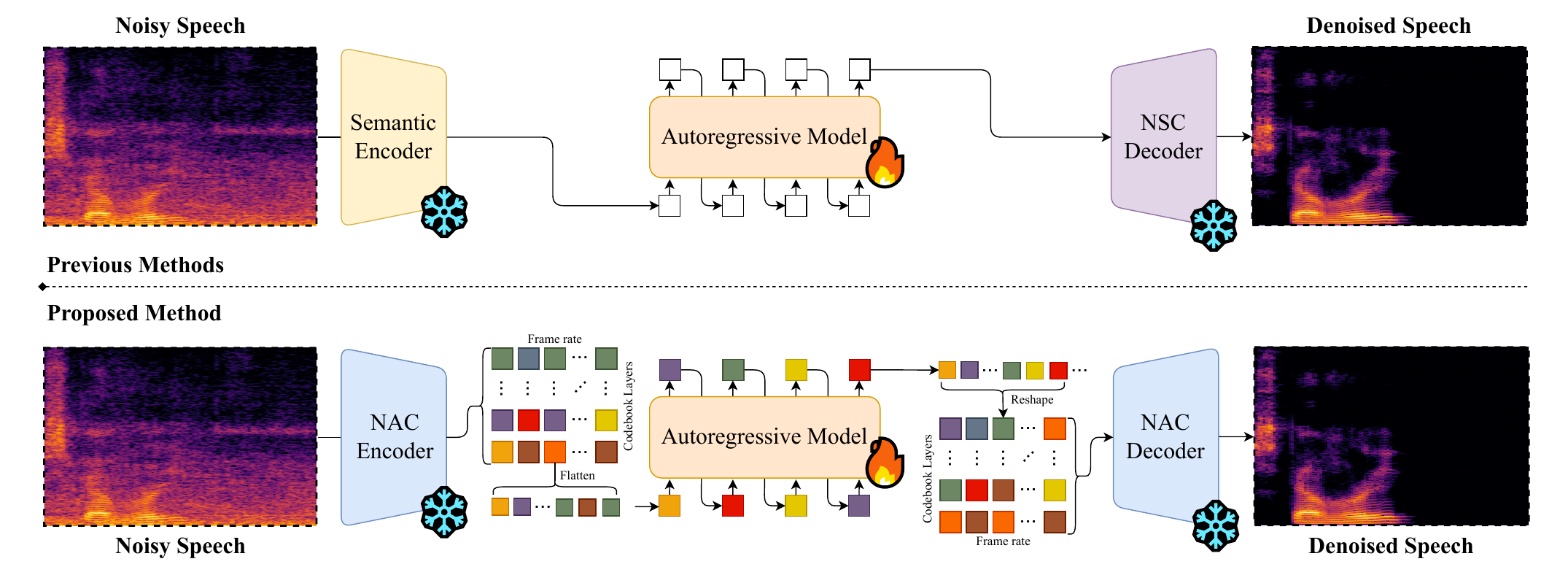}
    \caption{Overview of DAC-SE1 framework for high-fidelity speech enhancement and  bandwidth extension. Previous work mostly uses a continuous speech representation as the input to the autoregressive model (e.g., HuBERT or WavLM) and then predicts tokens from a Neural Speech Codec (NSC). These models are limited to 16 kHz signals. Our approach does not require semantic representations and only leverages the compressed representation of Neural Audio Codecs (NAC). We use the DAC model, compressing a 44.1 kHz signal into 9 codebook layers at 86 Hz framerate. We flatten this sequence into $9\cdot86$ tokens per second which are translated by our LlaMa-based model into clean speech in the DAC token space, which can then be reconstructed using the DAC decoder.}
    \label{fig:dac-se1}
\end{figure*}
To this end, \textbf{we introduce DAC-SE1, a simple LM-based speech enhancement framework} which uses discrete audio tokens to model high-resolution 44.1 kHz speech and audio signals. By leveraging high-fidelity audio tokens from DAC~\cite{kumar2023high} and scaling model capacity, DAC-SE1 performs both speech enhancement and bandwidth extension without auxiliary encoders, dual-channel conditioning, or multi-stage pipelines. We evaluate the performance of DAC-SE1 on widely used objective metrics and a MUSHRA human evaluation, demonstrating that, at sufficient scale, a single autoregressive LM can achieve high-fidelity SE and outperform previous state-of-the-art without requiring architectural modifications.

\medskip
\noindent In summary, our contributions are threefold:  
\begin{enumerate}  
    \item We propose DAC-SE1, a single-stage 1B parameter LM-based SE framework operating directly on 44.1 kHz DAC tokens, achieving high-fidelity speech enhancement and  bandwidth extension.\footnote{Samples available on https://lucala.github.io/dac-se1/} 
    \item We show empirically that scaling model size and training data allows the model to outperform prior LM-SE baselines, across both objective metrics and human evaluations, without requiring domain-specific adaptations.  
    \item We release our models and training pipeline to facilitate reproducibility and further research in scalable, high-quality speech enhancement.
\end{enumerate}

\section{Background}
\subsection{Time-Domain Speech Enhancement}
Traditional SE models operate directly in the time domain or in the time-frequency domain to map noisy inputs to clean outputs. Convolutional and recurrent architectures such as Conv-TasNet~\cite{luo2019conv}, Demucs~\cite{defossez2020real}, and DCCRN~\cite{hu2020dccrn} demonstrate good performance in time-domain enhancement and de-reverberation. However, these architectures are often tailored to specific distortions, require task-specific design, and do not naturally scale to high-fidelity or multi-task settings. Moreover, they do not leverage recent advances in transformer-based modeling nor integrate easily into multi-modal generative frameworks.

\subsection{Discrete Audio Representations}
Neural audio codecs such as EnCodec~\cite{defossez2022high}, X-Codec2~\cite{ye2025llasa}, and DAC~\cite{kumar2023high} learn to map a time-domain signal into compact sequences of discrete tokens via quantization schemes (e.g., vector quantization, residual vector quantization, or finite scalar quantization). In the case of residual vector quantization (RVQ), each frame of audio is represented by a stack of codebook entries: a coarse codebook captures the global signal structure, while subsequent residual codebooks refine the representation with increasingly fine acoustic details. This hierarchical structure allows codecs to balance compression efficiency with perceptual quality.
By operating on discrete token streams, speech models gain the advantage of bandwidth-efficient modeling with autoregressive transformers, which are able to capture long-range dependencies. While many neural codecs use RVQ, only a few have been scaled to very high perceptual quality at high-fidelity sampling rates (44.1 or 48 kHz). DAC~\cite{kumar2023high} is one such codec, providing discrete representations with fidelity close to uncompressed signals, which makes it particularly suitable for speech enhancement in high-fidelity settings.

\subsection{Language Models for Generative Audio}
Autoregressive transformers trained on discrete audio tokens have recently advanced a range of tasks, including text-to-speech (TTS)~\cite{ye2025llasa, borsos2023audiolm} and speech enhancement~\cite{kang2025llase, wang2024selm}. These methods demonstrate that LMs can jointly model long-range semantic structure and local acoustic detail. However, most current LM-based SE frameworks operate at 16 kHz resolution and rely on multi-stage or conditional pipelines (e.g., auxiliary encoders, noise estimators, or dual-channel modeling). Such constraints limit fidelity, increase complexity, and hinder scalability for unified speech enhancement models.

\section{Methodology}

\subsection{Discrete Audio Representation}
\label{sec:DAC}
We adopt the DAC codec~\cite{kumar2023high} at 44.1~kHz, which encodes audio into 9 residual codebooks, each containing a vocabulary size of 1024 codes. While some prior works preserve the multi-codebook structure by processing each codebook separately and later aggregating embeddings~\cite{li2024masksr}~\cite{li2024masksr}, we simplify the design by flattening all codebooks into a single time-major token sequence. This was shown by MusicGen~\cite{copet2023simple} to be a viable strategy when dealing with RVQ tokens. This strategy reduces architectural complexity and aligns with standard LM training pipelines, at the expense of longer token sequences. Since scaling laws indicate that larger causal LMs handle such longer contexts effectively, we find this simplification both practical and effective for high-resolution SE.
\subsection{Implementation Details}
Our core model is a causal transformer language model based on the 1B parameter LLaMA architecture~\cite{touvron2023llama}.
The model uses a hidden size of 1536, an intermediate feedforward size of 6144, 24 transformer layers, 24 attention heads (with 24 key-value heads), and a maximum sequence length of 8192 tokens, with all dimensions and depth scaled consistently for this parameter budget. 
To accommodate the long sequences resulting from flattened DAC tokens, we use rotary positional embeddings (RoPE)~\cite{su2024roformer} 
with a large scaling factor ($\theta=100{,}000$), which significantly improves stability and generalization to extended context lengths. Following insights from large language models, this design ensures that our model can capture both fine-grained acoustic structure and long-range token-structure dependencies.

\begin{table}[t]
\centering
\small
\begin{tabularx}{\columnwidth}{p{1.7cm} >{\centering\arraybackslash}p{0.7cm} X}
\toprule
\textbf{Distortion} & \textbf{Prob.} & \textbf{Hyperparameters} \\
\midrule
White Noise & 0.3 & SNR $\in [0, 25]$ dB \\
Noise & 0.7 & SNR $\in [-5, 20]$ dB \\
Reverb & 0.5 & -- \\
Downsampling & 0.5 & sr $\in \{2,4,6,8,16\}$ kHz \\
Packet loss & 0.3 & size $\in[50,200]$ ms,  $p_\text{drop} \in [0.02,0.2]$ \\
\bottomrule
\end{tabularx}
\caption{Distortion distribution in training dataset. Noise is added to clean speech, reverberation is simulated by convolving with RIRs, packet loss is applied by zeroing out affected segments, and downsampling is performed by reducing the sampling rate and resampling back to 44.1kHz.}
\label{tab:distortions}
\end{table}

\subsection{Training}
Training a general speech enhancement model involves multi-task optimization across a variety of distortions, including noise, reverberation, downsampling, and packet loss. A key challenge arises from the varying loss scales per task. For example, packet loss concealment exhibits a relatively low loss during training because most input tokens (noisy tokens) are the same as the corresponding clean tokens, while only a small fraction is different, namely the tokens corresponding to lost packets. As a result, the gradient contribution per task is uneven, which can cause joint training on all tasks to generalize poorly.
To address this, we adopt a two-stage training strategy. In the first stage, we perform standard multi-task training. In the second stage, we fine-tune the model per task, allowing each task to optimize its own loss more effectively. Importantly, this does not require separate models per task; the same model is iteratively fine-tuned on each task. We observe that this approach produces distinct and informative loss curves per task, leading to better generalization across all distortions.
Our model was trained on H200 GPUs for 12 hours on more than 5 billion tokens.

\subsection{Datasets}
For the reference clean speech, we use the HiFiTTS-2~\cite{langman2025hifitts} corpus, a high-quality 44.1 kHz speech dataset. From this corpus, we select a 2k-hour subset, truncating clips to a maximum of 5 seconds. For noise, we combine multiple open-source datasets to ensure diversity: MUSAN~\cite{snyder2015musan} (noise and music), DEMAND~\cite{thiemann2013diverse} (domestic and environmental recordings), Urban Acoustic Scenes~\cite{mesaros2018multi}, and WHAM!~\cite{wichern2019wham} noise. To simulate reverberation, we further include room impulse responses from the RIRS NOISES corpus, specifically OpenSLR 26 and 28~\cite{ko2017study}. 
We first generate our \emph{Stage-1} training dataset by following the distribution of distortions in \cref{tab:distortions}. Then, we generate the \emph{Stage-2} training datasets, which are task-specific, meaning each dataset corresponds to a unique label of distortion. All datasets are cleared of duplicates, i.e., samples that have the same clean speech. For faster training, we pre-process and encode the datasets using DAC and flattening to obtain a sequence of type: 

{\footnotesize
\[
\texttt{[Noisy DAC Tokens]} \;|\; \texttt{start-clean} \;|\; \texttt{[Clean DAC Tokens]}
\]
}

\noindent where \texttt{start-clean}  is a special boundary token marking the transition from the noisy signal to the clean signal.

\begin{table*}[t]
    \centering
    \begin{tabularx}{\linewidth}{%
        p{2.5cm} %
        >{\centering\arraybackslash}X %
        >{\centering\arraybackslash}X %
        >{\centering\arraybackslash}X %
        >{\centering\arraybackslash}X %
        >{\centering\arraybackslash}X %
        >{\centering\arraybackslash}p{1.7cm}       %
        >{\centering\arraybackslash}p{1.5cm}       %
        >{\centering\arraybackslash}X       %
        >{\centering\arraybackslash}p{1.5cm}  %
    }
        \toprule
        
        \textbf{Model} & OVRL$\uparrow$ & SIG$\uparrow$ & BAK$\uparrow$  & P808$\uparrow$ & PESQ$\uparrow$ & {S-BERTS$\uparrow$} & PLCMOS$\uparrow$ & WER $\downarrow$ & MUSHRA$\uparrow$\\
        \midrule
        Noisy           & 2.44 & 3.18 & 2.79 & 3.11 & 2.63 & 0.89  & 3.84 & 0.25 & 35.8 \\  
        Clean           & 3.03 & 3.41 & 3.80  & 3.64 & 4.50 & 1.00 & 4.41 & 0.00 & 94.5 \\
        \midrule\midrule
        LLaSE-G1        & 2.90 & 3.24 & 3.83 & 3.47 & 1.98 & 0.86 & 4.19 & 0.27 & 44.1 \\
        VoiceFixer       & 2.92 & 3.21 & \textbf{3.90} & 3.43 & 1.85 & 0.81  & 4.29 & 0.45 & 34.5 \\
        \textbf{DAC-SE1 (ours)}   & \textbf{2.95} & \textbf{3.33} & 3.70 & \textbf{3.56} & \textbf{2.46} & \textbf{0.89} & \textbf{4.35} & \textbf{0.25} & \textbf{58.3} \\
        \bottomrule
    \end{tabularx}
    \caption{Comparison of LLaSE-G1~\cite{kang2025llase}, VoiceFixer~\cite{liu2021voicefixer}, and our model on the HiFiTTS-2 test set. 
    Objective metrics include DNSMOS OVRL/SIG/BAK~\cite{reddy2021dnsmosnonintrusiveperceptualobjective}, P.808~\cite{naderi2020an}, PESQ~\cite{PESQ}, SpeechBERTScore (S-BERTS)~\cite{saeki2024speechbertscorereferenceawareautomaticevaluation}, PLCMOS~\cite{diener2023plcmosdatadrivennonintrusive}, and WER computed using Whisper-Large~\cite{radford2022robustspeechrecognitionlargescale}. Subjective evaluation is done with MUSHRA. DAC-SE1 consistently outperforms prior systems in both objective and human evaluation.}

    \label{tab:subjective}
\end{table*}

\section{Evaluation}
We evaluate our models on speech enhancement datasets using widely adopted metrics. Specifically, we test on the ICASSP 2022 Packet Loss Concealment (PLC) challenge~\cite{diener2022interspeech} and the ICASSP 2023 DNS-challenge~\cite{dubey2023icassp2023deepnoise}. Additionally, we compare our models against other baselines on a small test set randomly sampled from HiFiTTS-2, DEMAND, and RIRS NOISES. This dataset is used for both objective and subjective evaluations and is disjoint from the training set in terms of both speakers and noise sources.

\smallskip
\noindent \textbf{Subjective Evaluation.}
We conduct a MUSHRA listening test, the gold-standard for estimating the quality of speech enhancement.\footnote{MUSHRA was conducted on https://www.mabyduck.com} The study included 26 participants. Each participant completed 12 trials, where the first two trials were training runs to familiarize the participants with the tool and task. All participants used headphones and were asked to be in a quiet environment. Each trial consisted of a clean reference signal, the degraded signal (used as a low anchor), a hidden reference, and the reconstructions of the models.

\subsection{Results}

\smallskip
\noindent \textbf{HiFiTTS-2 Evaluation.}
Table~\ref{tab:subjective} shows the objective evaluation results on the HiFiTTS-2 test set. Our model consistently outperforms both LLaSE-G1~\cite{kang2025llase} and VoiceFixer~\cite{liu2021voicefixer} across the majority of metrics, achieving stronger overall quality, speech naturalness, and perceptual consistency. Notably, while VoiceFixer performs slightly better in background suppression (BAK), our approach provides a more balanced improvement across all dimensions, leading to the best overall performance. The MUSHRA listening test further supports these findings. Human listeners consistently preferred the outputs of our model over both LLaSE-G1 and VoiceFixer.

\smallskip
\noindent \textbf{SE Benchmarks.}
On the ICASSP PLC challenge (see~\cref{tab:PLC-challenge}), our method achieves state-of-the-art perceptual quality, surpassing prior baselines in PLCMOS while remaining competitive in overall quality. On the DNS challenge, our approach performs on par with strong published baselines across widely adopted perceptual metrics, as shown in \cref{tab:DNS-Challenge}, confirming that the model generalizes effectively beyond the custom training data and adapts to previously unseen profiles of noise.

\begin{table}[t]
    \centering
    \begin{tabularx}{\columnwidth}{p{3cm}
    >{\centering\arraybackslash}X
    >{\centering\arraybackslash}X
    >{\centering\arraybackslash}X}
        \toprule
        Model & OVRL$\uparrow$ & PLCMOS$\uparrow$ \\
        \midrule
        Noisy & 2.56 & 2.90 \\
        LPCNet~\cite{valin2019lpcnetimprovingneuralspeech} & 3.09 & 3.74 \\
        BS-PLCNet~\cite{zhang2024bsplcnetbandsplitpacketloss} & 3.20 & 4.29 \\
        LLaSE-G1 single  & 3.03 & 3.68 \\
        LLaSE-G1 multi   & \textbf{3.27} & 4.30 \\
        DAC-SE1 (ours)  & 3.12 & \textbf{4.34} \\  %
        \bottomrule
    \end{tabularx}
    \caption{DNSMOS OVRL and PLCMOS scores on ICASSP 2022 PLC-challenge blind testset.}
    \label{tab:PLC-challenge}
\end{table}

\begin{table}[t]
    \centering
    \begin{tabularx}{\columnwidth}{p{3cm}
    >{\centering\arraybackslash}X
    >{\centering\arraybackslash}X
    >{\centering\arraybackslash}X
    >{\centering\arraybackslash}X}
        \toprule
        Model & SIG$\uparrow$ & BAK$\uparrow$ & OVRL$\uparrow$ \\
        \midrule
        Noisy & 4.15 & 2.37 & 2.71 \\
        TEA-PSE 3.0~\cite{ju2023teapse30tencentetherealaudiolabpersonalized} & 4.12 & \textbf{4.05} & 3.65 \\
        NAPSE~\cite{yan2023npuelevocpersonalizedspeechenhancement} & 3.81 & 3.99 & 3.38 \\
        LLaSE-G1 single & \textbf{4.21} & 3.99 & \textbf{3.72} \\
        LLaSE-G1 multi &  4.20 & 3.97 & 3.70 \\

        DAC-SE1 (ours)   & 4.18 & 3.80 & 3.63 \\  %
        \bottomrule
    \end{tabularx}
    \caption{pDNSMOS scores on ICASSP 2023 DNS-challenge blind testset.}
    \label{tab:DNS-Challenge}
\end{table}

\section{Conclusion}
We introduced DAC-SE1, an LM-based SE framework that operates directly on DAC tokens, achieving high-fidelity speech enhancement without auxiliary encoders or multi-stage pipelines. Experiments demonstrate that DAC-SE1 outperforms prior LM-based SE methods across objective metrics and in human evaluations. Our results show that speech enhancement methods benefit from scaling laws, a trend we expect will shape the next generation of SE models.

\ninept
\bibliographystyle{IEEEbib}
\bibliography{references}

\end{document}